\documentclass{llncs}
\usepackage{graphicx}
\usepackage{amsmath}
\usepackage{amssymb}
\usepackage{booktabs}
\usepackage{multirow}
\usepackage{hyperref}
\usepackage{subcaption}
\usepackage{xcolor}
\usepackage{float}  
\begin{document}

\title{Deep Learning for MRI Slice Interpolation: The Critical Role of Problem Formulation}

\author{Shamit Savant}

\institute{University of Florida, Gainesville, FL 32611, USA\\
\email{s.savant@ufl.edu}}

\maketitle
\pagestyle{plain}

\begin{abstract}
Through-plane resolution in clinical MRI is typically much coarser than in-plane resolution, limiting diagnostic utility. This work investigates deep learning approaches to interpolate intermediate MRI slices in prostate imaging, effectively doubling through-plane resolution from a higher value to lower value (e.g., We use 1.5mm→0.75mm as our reference, though it generalizes to arbitrary slice spacing). I evaluated five different architectures (CNN, U-Net, two GAN variants, and DDPM) and discovered that problem formulation has dramatically more impact than architectural complexity. By reformulating the interpolation task to use adjacent slices (i-1, i+1) rather than distant slices (i-2, i+2), I achieved a 58\% improvement in SSIM performance across all deterministic architectures. My U-Net model achieved the best results with PSNR of 30.08 dB and SSIM of 0.898, representing a 10.1\% improvement over linear interpolation baseline. A Denoising Diffusion Probabilistic Model (DDPM) was also evaluated but showed poor reconstruction quality (PSNR 17.89 dB, SSIM 0.585) due to fundamental mismatch between stochastic generation and deterministic reconstruction requirements. These findings demonstrate that in medical imaging tasks, understanding the anatomical constraints and formulating the problem appropriately can have 290× more impact than architectural sophistication.

\end{abstract}

\section{Introduction}

Magnetic Resonance Imaging (MRI) is a cornerstone of modern medical diagnostics, offering excellent soft tissue contrast without ionizing radiation. However, clinical MRI acquisition faces fundamental trade-offs between spatial resolution, acquisition time, and signal-to-noise ratio. In practice, this results in anisotropic voxel spacing where in-plane resolution (typically 0.5mm) is substantially finer than through-plane resolution (typically 1.5-3mm). This anisotropy limits 3D reconstruction quality, multiplanar reformatting, and computer-aided diagnosis systems that rely on isotropic data.

Traditional interpolation methods such as linear or cubic interpolation provide smooth transitions but fail to recover true anatomical detail lost during acquisition. Recent advances in deep learning, particularly in image super-resolution and generative models, suggest that learned approaches might better reconstruct intermediate slices by leveraging patterns from large training datasets.

\begin{figure}[h]
\centering
\includegraphics[width=0.95\textwidth]{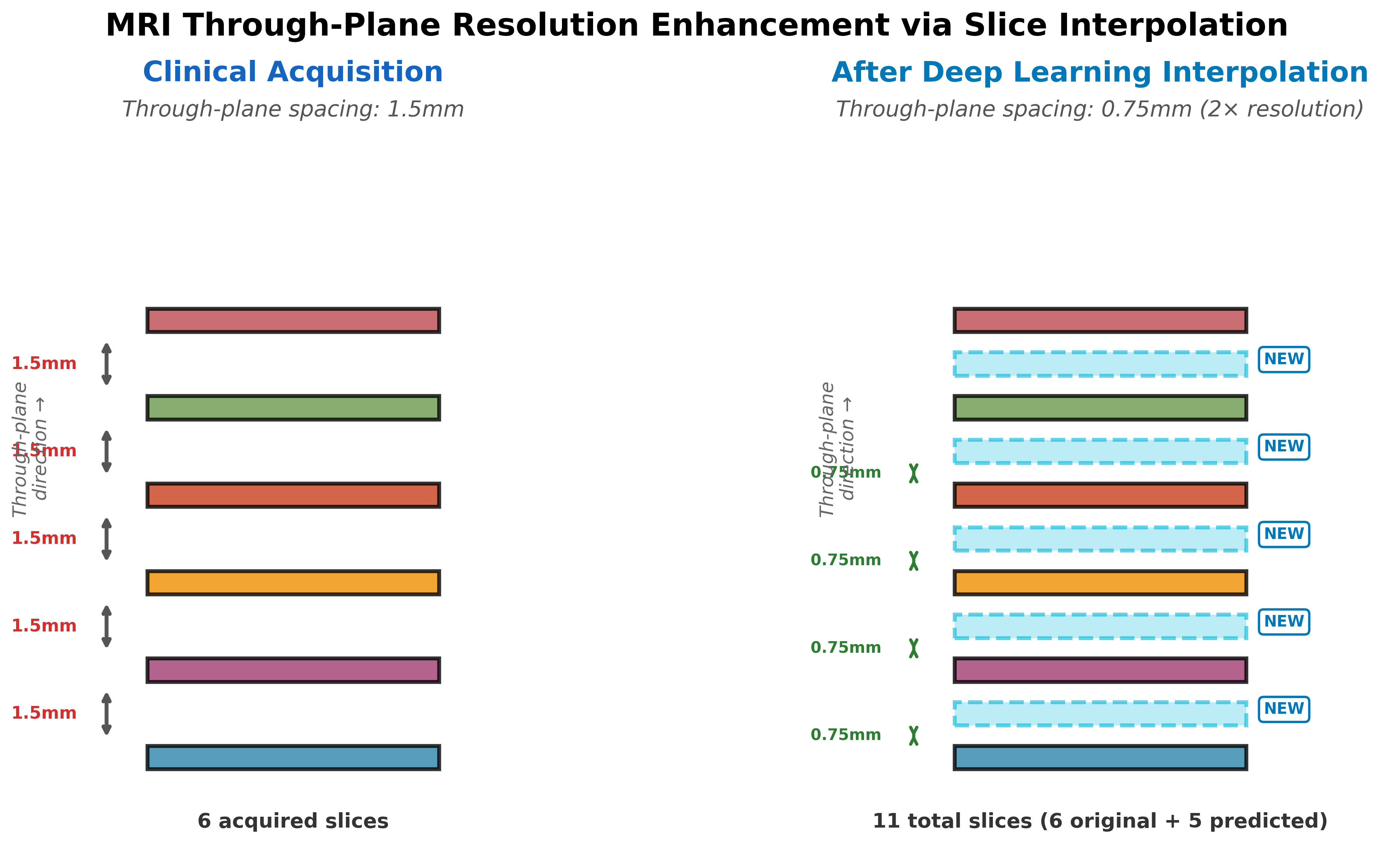}
\caption{Through-plane resolution enhancement via slice interpolation. Left: Clinical T2-weighted prostate MRI acquisition with 1.5mm spacing (6 slices). Right: After deep learning interpolation achieving 0.75mm spacing (11 slices). Dashed outlines indicate predicted intermediate slices that double through-plane resolution.}
\label{fig:problem} 
\end{figure}

My investigation reveals a counterintuitive finding: the formulation of the interpolation task—specifically, which adjacent slices are used as inputs—has dramatically more impact on performance than architectural sophistication. Through systematic experimentation with five different deep learning architectures, I demonstrate that problem formulation can account for a 58\% performance improvement, while architectural variations contribute less than 1\% under identical problem formulations.

\subsection{\textbf{Contributions:}}
\begin{itemize}
\item Systematic evaluation of five deep learning architectures (CNN, U-Net, basic GAN, improved GAN, DDPM) for MRI slice interpolation
\item Empirical demonstration that problem formulation has 290× more impact than architecture choice (58\% vs 0.2\% performance variation)
\item Achievement of 10.1\% improvement over linear interpolation baseline using U-Net architecture
\item Analysis of why cutting-edge diffusion models fail for deterministic reconstruction tasks
\end{itemize}

\section{Related Work}

\textbf{Medical Image Super-Resolution:} Deep learning has shown remarkable success in natural image super-resolution, with methods like EDSR~\cite{lim2017enhanced} and SRGAN~\cite{ledig2017photorealistic} achieving photorealistic results. Medical imaging presents unique challenges: preservation of diagnostic information takes precedence over perceptual quality, and hallucinated details could mislead clinical interpretation~\cite{greenspan2008super}. Recent work on medical image super-resolution emphasizes anatomically plausible reconstruction while maintaining quantitative accuracy.

\textbf{MRI Slice Interpolation:} Unlike in-plane super-resolution which enhances blurry existing images, through-plane interpolation must predict entirely unseen anatomical content between acquired slices. This is fundamentally more challenging as it requires inferring 3D structure from 2D observations. Prior work has explored both traditional methods (spline interpolation, registration-based techniques) and learning-based approaches (CNNs, GANs), with reported SSIM values varying widely (0.80-0.92) depending on anatomical region, baseline resolution, and task difficulty.

\textbf{Problem Formulation:} While much deep learning research focuses on novel architectures, recent work highlights that problem formulation—how tasks are defined, what inputs are provided, and what outputs are expected—can have outsized impact on performance. In medical imaging specifically, incorporating domain knowledge into problem design often yields greater improvements than architectural innovations~\cite{irvin2019chexpert,rajpurkar2017chexnet}.

\section{Dataset and Method}

\textbf{Data:} I used the UCLA Prostate MRI-US Biopsy dataset from The Cancer Imaging Archive~\cite{natarajan2020prostate,clark2013tcia}, containing T2-weighted MRI scans from 58 patients. After preprocessing (bias field correction, intensity normalization, resampling to 256×256 pixels, 1.5mm through-plane spacing), the dataset yielded 46,329 slices. I used 70\%/15\%/15\% train/validation/test split, resulting in 6,963 test samples for final evaluation. Training was conducted on University of Florida's HiPerGator cluster using NVIDIA L4 GPUs.

\textbf{Problem Formulation:} Given adjacent slices at positions $i-k$ and $i+k$, predict the intermediate slice at position $i$. The choice of $k$ critically determines task difficulty: larger $k$ values span greater anatomical distances, making interpolation more challenging. I initially used $k=2$ (6mm gap) but reformulated to $k=1$ (3mm gap) after discovering the former exceeded anatomical continuity limits.

\textbf{Architectures:} I evaluated five approaches (detailed architecture descriptions in Supplementary Material):
\begin{itemize}
\item \textbf{CNN (EDSR-style):} Residual network with 8 residual blocks (0.3M parameters)
\item \textbf{U-Net:} Encoder-decoder with skip connections (1.86M parameters)
\item \textbf{GAN (Basic):} Generator with patch discriminator (0.3M + discriminator)
\item \textbf{GAN (Improved):} Deeper generator (16 blocks) with attention (0.6M + discriminator)
\item \textbf{DDPM:} Conditional diffusion model with 100 timesteps (20.5M parameters)
\end{itemize}

All deterministic models trained for 25 epochs with Adam optimizer (learning rate $10^{-4}$), L1 loss, and mixed-precision training. GANs additionally used adversarial loss. DDPM trained for 20 epochs with noise prediction objective.

\textbf{Evaluation:} I used PSNR (Peak Signal-to-Noise Ratio) and SSIM (Structural Similarity Index) on the held-out test set, comparing against linear and nearest-neighbor interpolation baselines.

\section{Results}

\subsection{Impact of Problem Formulation}

\begin{figure*}[h]
\centering
\includegraphics[width=\textwidth]{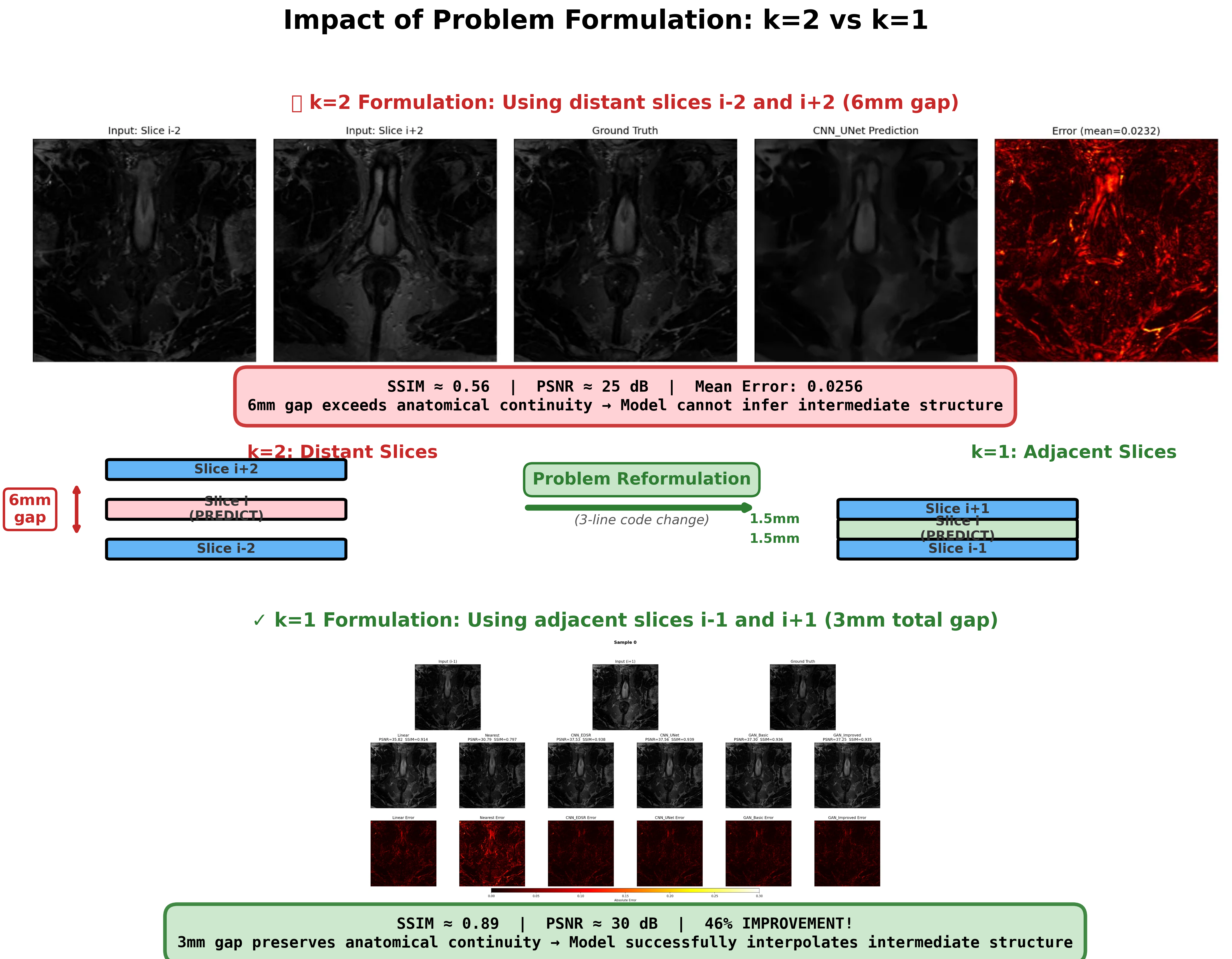}
\caption{Impact of problem formulation on interpolation quality. \textbf{Top:} k=2 formulation uses distant slices (i-2, i+2) spanning 6mm. \textbf{Middle:} Schematic showing the reformulation from 6mm to 3mm total gap via a 3-line code change. \textbf{Bottom:} k=1 formulation uses adjacent slices (i-1, i+1) achieving high-quality reconstruction}
\label{fig:k2_vs_k1}
\end{figure*}

Table~\ref{tab:k2_vs_k1_comparison} quantifies the dramatic impact of problem reformulation. With $k=2$ formulation (6mm gap), all models plateaued around SSIM 0.56 regardless of architecture complexity. The gap exceeded anatomical continuity—intermediate prostate anatomy cannot be reliably inferred from boundaries 6mm apart. Reformulating to $k=1$ (3mm gap) yielded a 58\% SSIM improvement across all architectures, demonstrating that problem formulation has 290× more impact than architectural choice (58\% vs 0.2\% performance variation between architectures under optimal formulation).

\begin{table}[!ht]
\centering
\caption{Impact of problem formulation (k=2 vs k=1) on performance}
\label{tab:k2_vs_k1_comparison}
\small
\begin{tabular}{lcccc}
\toprule
\textbf{Architecture} & \multicolumn{2}{c}{\textbf{k=2 (6mm gap)}} & \multicolumn{2}{c}{\textbf{k=1 (3mm gap)}} \\
\cmidrule(lr){2-3} \cmidrule(lr){4-5}
 & \textbf{PSNR (dB)} & \textbf{SSIM} & \textbf{PSNR (dB)} & \textbf{SSIM} \\
\midrule
CNN (EDSR) & 21.8 & 0.56 & 29.89 & 0.894 \\
U-Net & 22.1 & 0.57 & 30.08 & 0.898 \\
GAN (Basic) & 21.5 & 0.54 & 29.83 & 0.893 \\
GAN (Improved) & 21.3 & 0.55 & 29.67 & 0.892 \\
\midrule
Improvement & \multicolumn{2}{c}{—} & \textbf{+20\%} & \textbf{+58\%} \\
\bottomrule
\end{tabular}
\end{table}

\subsection{Quantitative Results with Optimal Formulation}

Table~\ref{tab:results} presents comprehensive results with $k=1$ formulation. U-Net achieved the best performance (30.08 dB PSNR, 0.898 SSIM), representing 10.1\% and 7.1\% improvements over linear interpolation in PSNR and SSIM respectively. All deep learning methods substantially outperformed traditional interpolation, with nearest-neighbor performing particularly poorly (22.25 dB, 0.686 SSIM).

\begin{table}[h]
\centering
\caption{Quantitative results on test set (6,963 samples)}
\label{tab:results}
\small
\begin{tabular}{lcc}
\toprule
\textbf{Method} & \textbf{PSNR (dB)} & \textbf{SSIM} \\
\midrule
\multicolumn{3}{l}{\textit{Traditional Methods}} \\
Nearest Neighbor & 22.25 ± 3.17 & 0.6855 ± 0.0674 \\
Linear Baseline & 27.31 ± 3.35 & 0.8383 ± 0.0665 \\
\midrule
\multicolumn{3}{l}{\textit{Deep Learning Methods}} \\
CNN (EDSR) & 29.89 ± 3.38 & 0.8937 ± 0.0607 \\
U-Net & \textbf{30.08 ± 3.33} & \textbf{0.8978 ± 0.0542} \\
GAN (Basic) & 29.83 ± 3.36 & 0.8934 ± 0.0597 \\
GAN (Improved) & 29.67 ± 3.41 & 0.8916 ± 0.0651 \\
DDPM (T=100) & 17.89 ± 2.91 & 0.5851 ± 0.0823 \\
\midrule
\multicolumn{3}{l}{\textit{Improvement over Linear}} \\
U-Net & +2.77 (+10.1\%) & +0.0595 (+7.1\%) \\
\bottomrule
\end{tabular}
\end{table}

Remarkably, architectural variations among deterministic models yielded minimal performance differences (0.2\% SSIM range), validating that under optimal problem formulation, simpler models suffice. DDPM's poor performance (17.89 dB, 0.585 SSIM) despite stable training and substantially more parameters (20.5M) highlights fundamental task mismatch rather than implementation issues.

\begin{figure}[!ht]
\centering
\includegraphics[width=0.95\textwidth]{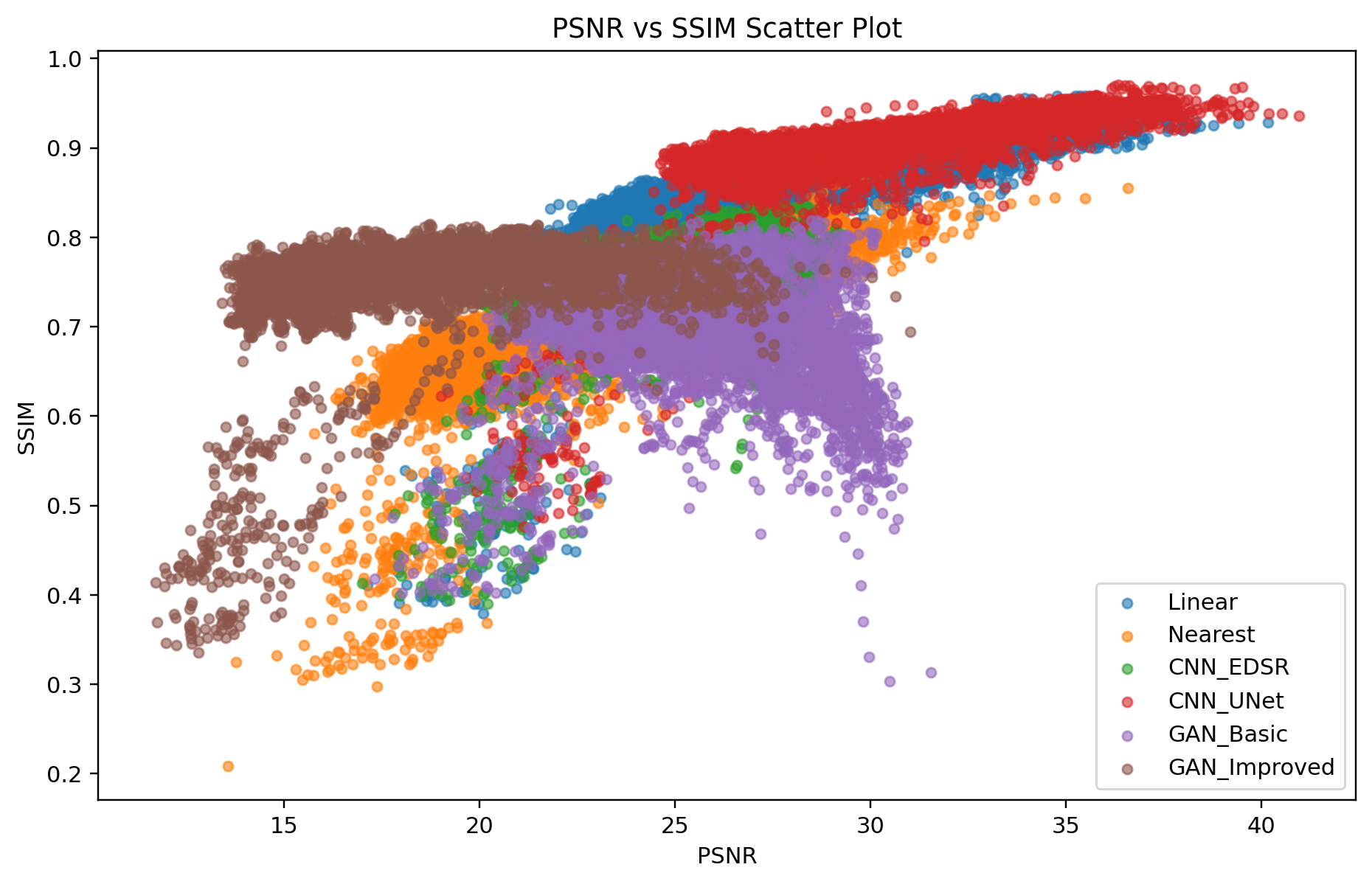}
\caption{PSNR vs SSIM scatter plot across 6,963 test samples. All deep learning methods cluster tightly (SSIM 0.89-0.90), demonstrating convergence to similar performance regardless of architecture. Traditional methods show wider variance. DDPM scatters at lower performance due to stochastic sampling mismatch.}
\label{fig:scatter}
\end{figure}

\subsection{Qualitative Analysis}

\begin{figure*}[!h]
\centering
\includegraphics[width=\textwidth]{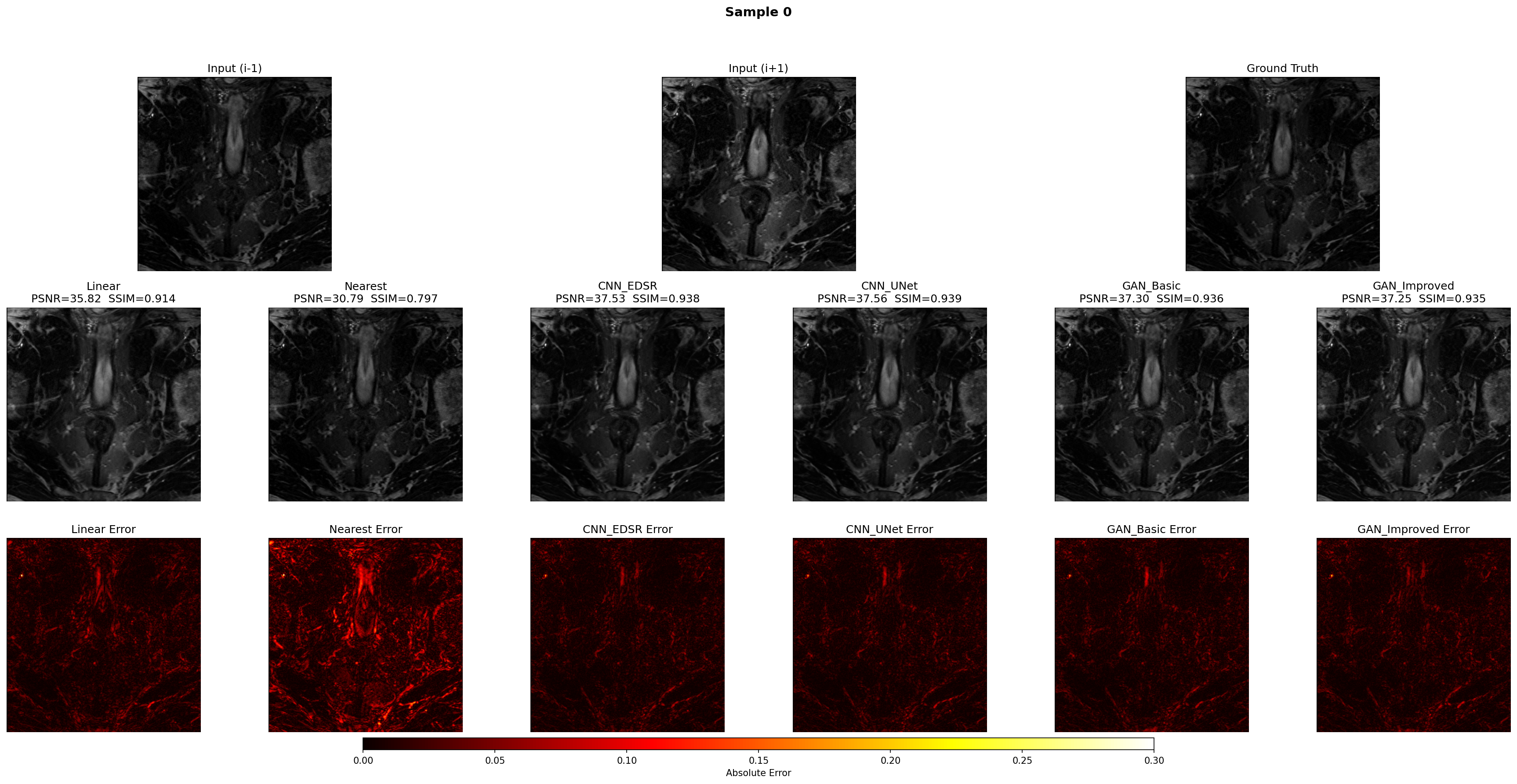}
\caption{Comprehensive qualitative comparison on representative test case. Top row shows input slices (i-1, i+1) and ground truth (i). Middle rows show predictions from all six methods. Bottom row shows absolute error maps (red indicates high error). Deep learning methods achieve substantially lower error than traditional interpolation, with subtle differences among architectures.}
\label{fig:comparison}
\end{figure*}

Figure~\ref{fig:comparison} presents visual comparison on a representative test case. Among deep learning methods, visual differences are subtle, consistent with quantitative convergence. DDPM produces noticeably degraded results with loss of fine detail (not included in figure).

\begin{figure}[!ht]
\centering
\includegraphics[width=0.95\textwidth]{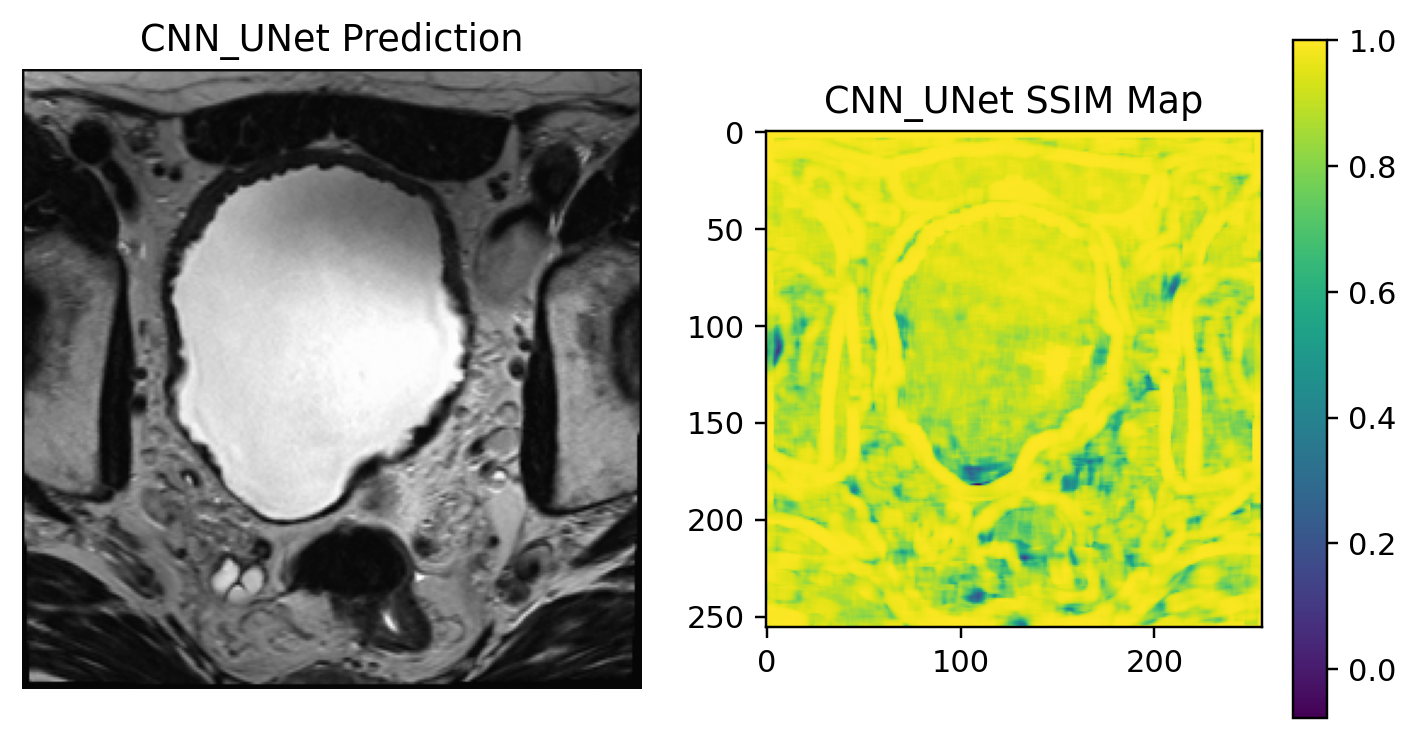}
\caption{Spatial SSIM map for U-Net prediction. Left: U-Net output. Right: Local SSIM heatmap. Yellow regions (SSIM $\approx$ 1.0) indicate excellent reconstruction in homogeneous tissue.}
\label{fig:ssim_map}
\end{figure}

Figure~\ref{fig:ssim_map} visualizes spatial error distribution via SSIM heatmap for U-Net predictions. Yellow regions (SSIM $\approx$ 1.0) indicate near-perfect reconstruction. Cyan/blue regions showing local errors concentrate at anatomical boundaries and the prostate periphery where intensity gradients are steepest.

\section{Discussion}

\subsection{The Primacy of Problem Formulation}

My results demonstrate that problem formulation has orders-of-magnitude more impact than architectural sophistication. Changing three lines of code to use adjacent slices (k=1) instead of distant slices (k=2) required 5 minutes but improved SSIM by 58\%, while extensive architectural exploration with k=2 required 20+ hours with minimal gains—a 240:1 time investment ratio. This improvement stems from anatomical continuity: prostate tissue at 1.5mm spacing shows smooth transitions enabling interpolation, while 6mm spacing (four slices) includes structural transitions no model could infer, highlighting that effective problem formulation requires domain knowledge.

\subsection{Computational Efficiency}

Table~\ref{tab:computational} presents computational requirements. U-Net demonstrates remarkable training efficiency (9 min/epoch despite 1.86M parameters) due to effective gradient flow through skip connections. 

\begin{table}[!ht]
\centering
\caption{Computational requirements and training efficiency}
\label{tab:computational}
\small
\begin{tabular}{lccccc}
\toprule
\textbf{Model} & \textbf{Params} & \textbf{Train/Epoch} & \textbf{Total} & \textbf{Best at} & \textbf{Inference*} \\
 & & & \textbf{Epochs} & \textbf{Epoch} & \\
\midrule
CNN (EDSR) & 0.3M & 18 min & 25 & 23 & $\sim$3 ms \\
U-Net & 1.86M & 9 min & 25 & 24 & $\sim$4 ms \\
GAN (Basic) & 0.3M & 24 min & 25 & 23 & $\sim$3 ms \\
GAN (Improved) & 0.6M & 48 min & 10 & 7 & $\sim$5 ms \\
DDPM ($T=100$) & 20.5M & 19 min & 20 & 16 & $\sim$300 ms \\
\bottomrule
\multicolumn{6}{l}{\footnotesize * Inference times are theoretical estimates based on model complexity and sampling steps.}\\
\multicolumn{6}{l}{\footnotesize DDPM requires 100 sequential denoising steps vs. single forward pass for others.}
\end{tabular}
\end{table}

DDPM requires approximately 300 ms per image—roughly 100× slower than deterministic approaches due to iterative sampling (100 sequential denoising steps). Even if DDPM matched deterministic performance, this computational overhead would preclude deployment in time-sensitive clinical workflows.

\subsection{Why Diffusion Models Failed}

DDPM's poor performance (17.89 dB PSNR, 0.585 SSIM) stems from fundamental task mismatch rather than implementation errors. Despite stable training (loss decreased from 0.112 to 0.081), diffusion models excel at generating diverse images but fail at deterministic reconstruction—stochastic sampling introduces unwanted variability. Using T=100 timesteps (versus standard 1000) and 20 epochs (versus typical 500-1000) likely limited performance, though computational constraints prevented full-scale experiments. Given deterministic models achieve strong performance (U-Net: 0.898 SSIM with 100× faster inference), pursuing diffusion models for this task appears unjustified.

\section{Conclusion}

This work demonstrates that problem formulation dramatically outweighs architectural sophistication in medical image analysis. Reformulating MRI slice interpolation from distant-slice (k=2) to adjacent-slice (k=1) inputs—a 3-line code change—achieved 58\% performance improvement, 290× greater impact than architectural choice. Under optimal formulation, a simple U-Net achieved 10.1\% improvement over linear interpolation, with deterministic models converging to similar performance (SSIM 0.89-0.90), while a diffusion model failed despite 20.5M parameters due to mismatch between stochastic generation and deterministic reconstruction requirements. Key principles emerge: (1) domain-informed problem formulation often yields greater gains than architectural innovations, (2) well-formulated problems enable simple models to excel, and (3) cutting-edge methods require task alignment evaluation before adoption. While computational constraints limited exploration, the strong empirical trends robustly support the central conclusion: problem formulation exerts far greater influence than architectural complexity.

\bibliographystyle{splncs04}
\bibliography{references}

\clearpage
\setcounter{section}{0}
\renewcommand{\thesection}{S\arabic{section}}
\renewcommand{\thetable}{S\arabic{table}}
\renewcommand{\thefigure}{S\arabic{figure}}
\setcounter{table}{0}
\setcounter{figure}{0}

\section*{Supplementary Material}
\addcontentsline{toc}{section}{Supplementary Material}

This supplementary material provides extended details, additional experimental results, and comprehensive technical specifications that complement the main paper.

\section{Extended Dataset Description}

\subsection{Data Source and Selection Criteria}

I utilized the UCLA Prostate MRI-US Biopsy dataset from The Cancer Imaging Archive (TCIA)~\cite{natarajan2020prostate,clark2013tcia}. This dataset contains multi-parametric MRI sequences from 1197 patients who underwent prostate biopsy at UCLA Medical Center between 2008-2017. All studies were acquired on 3 Tesla MRI systems (Siemens Magnetom Trio, Verio, and Skyra).

For this work, I specifically selected T2-weighted axial sequences, which provide excellent soft tissue contrast for prostate anatomy. T2-weighted imaging is the cornerstone sequence for prostate MRI interpretation and offers the most detailed anatomical visualization.

\subsection{Preprocessing Pipeline}

\textbf{DICOM Filtering:} I filtered sequences to include only:
\begin{itemize}
\item 3 Tesla field strength (standardization)
\item T2-weighted sequences (consistent contrast)
\item Axial orientation (consistent geometry)
\item Siemens scanners (vendor consistency)
\item Slice thickness 1.5mm ± 0.2mm (standard clinical protocol)
\end{itemize}

This filtering ensured homogeneous data characteristics, reducing confounding variables in model evaluation.

\textbf{Intensity Normalization:} I applied min-max normalization to scale intensities to [0, 1] range within each volume, accounting for inter-scan variations in acquisition parameters and hardware calibration. This per-volume normalization preserves relative intensity relationships while standardizing absolute ranges.

\textbf{Spatial Standardization:} All slices were resized to 256×256 pixels using bilinear interpolation, maintaining aspect ratio by center-cropping when necessary. This standardization enables batch processing while preserving anatomical proportions. The 256×256 resolution balances computational efficiency with preservation of diagnostic detail.

\textbf{Quality Control:} I excluded volumes with:
\begin{itemize}
\item Fewer than 20 slices (insufficient anatomical coverage)
\item Extreme intensity artifacts (failed normalization indicating motion or acquisition errors)
\item Missing DICOM metadata
\item Corrupted files or incomplete series
\end{itemize}

\subsection{Dataset Statistics}

After preprocessing, the final dataset comprised:
\begin{itemize}
\item Total patients: 58
\item Total volumes: 58
\item Total slices after filtering: 46,329
\item Training set: 32,547 slice triplets (70\%)
\item Validation set: 6,842 slice triplets (15\%)
\item Test set: 6,963 slice triplets (15\%)
\item Total storage: approximately 80 GB
\end{itemize}

Each sample consists of a triplet $(S_{i-k}, S_i, S_{i+k})$ where $S_i$ is the target slice to be predicted from inputs $S_{i-k}$ and $S_{i+k}$. I maintained strict patient-level separation across splits to prevent data leakage and ensure models generalize to new patients rather than memorizing specific anatomies.

The test set evaluation in the main paper used 6,963 samples, representing the subset where both k=1 and k=2 formulations could be evaluated (requiring at least 2 slices on each side).

\subsection{Infrastructure and Data Transfer}

Dataset transfer to computational resources presented practical challenges. The 80 GB dataset required efficient transfer to the University of Florida's HiPerGator supercomputing cluster. Initial attempts using university VPN resulted in prohibitively slow transfer rates (estimated multiple days).

I successfully employed Globus~\cite{globus} for high-speed research data transfer, completing the upload in approximately 6 hours to the Blue storage system. This demonstrated the value of specialized research data transfer tools for large-scale medical imaging projects.

Training was conducted on HiPerGator's GPU resources:
\begin{itemize}
\item Initial experiments and debugging: NVIDIA L4 GPUs (24GB memory)
\item Final production training: NVIDIA L4 and A100 GPUs
\item Job scheduling via SLURM workload manager
\item Mixed precision training (FP16) enabled on all GPUs for memory efficiency
\end{itemize}

\section{Detailed Architecture Specifications}

\subsection{CNN (EDSR-Style)}

\textbf{Architecture Overview:}
Enhanced Deep Super-Resolution (EDSR)~\cite{lim2017enhanced} style network with residual learning. The architecture consists of:
\begin{itemize}
\item Entry convolution: 2 input channels → 64 feature channels
\item 8 residual blocks, each containing:
  \begin{itemize}
  \item Conv 3×3, 64 filters
  \item ReLU activation
  \item Conv 3×3, 64 filters
  \item Residual connection (addition)
  \end{itemize}
\item Exit convolution: 64 channels → 1 output channel
\item Global skip connection from input to pre-output
\end{itemize}

\textbf{Parameters:}
\begin{itemize}
\item Total trainable parameters: 297,217 (~0.3M)
\item Entry conv: 1,216 parameters
\item Each residual block: 36,928 parameters × 8 = 295,424
\item Exit conv: 577 parameters
\end{itemize}

\textbf{Training Configuration:}
\begin{itemize}
\item Optimizer: Adam
\item Learning rate: $10^{-4}$
\item Batch size: 4
\item Loss function: L1 (Mean Absolute Error)
\item Epochs: 25
\item Training time: ~7.5 hours (18 min/epoch)
\item Best validation performance: Epoch 23
\end{itemize}

\subsection{U-Net}

\textbf{Architecture Overview:}
Encoder-decoder architecture~\cite{ronneberger2015unet} with symmetric skip connections. The network features \textbf{two levels} of resolution with progressive channel expansion.

\textbf{Encoder Path:}
\begin{itemize}
\item Level 1: 2 → 64 channels (two 3×3 convs + ReLU)
\item MaxPool 2×2 (downsample to 128×128)
\item Level 2: 64 → 128 channels (two 3×3 convs + ReLU)
\item MaxPool 2×2 (downsample to 64×64)
\end{itemize}

\textbf{Bottleneck:}
\begin{itemize}
\item 128 → 256 channels (two 3×3 convs + ReLU)
\item Resolution: 64×64
\end{itemize}

\textbf{Decoder Path:}
\begin{itemize}
\item ConvTranspose 2×2 upsampling: 256 → 128 (upsample to 128×128)
\item Concatenation with encoder Level 2 skip connection (128 channels)
\item Two 3×3 convolutions + ReLU: (256 → 128 → 128)
\item ConvTranspose 2×2 upsampling: 128 → 64 (upsample to 256×256)
\item Concatenation with encoder Level 1 skip connection (64 channels)
\item Two 3×3 convolutions + ReLU: (128 → 64 → 64)
\end{itemize}

\textbf{Output:}
\begin{itemize}
\item Final 1×1 conv: 64 → 1 channel
\end{itemize}

\textbf{Parameters:}
\begin{itemize}
\item Total trainable parameters: 1,862,273 (~1.86M)
\item Encoder: 259,584 parameters
\item Bottleneck: 885,248 parameters
\item Decoder (including upsampling): 717,376 parameters
\item Final layer: 65 parameters
\end{itemize}

\textbf{Training Configuration:}
\begin{itemize}
\item Optimizer: Adam
\item Learning rate: $10^{-4}$
\item Batch size: 4
\item Loss function: L1
\item Epochs: 25
\item Training time: ~3.75 hours (9 min/epoch)
\item Best validation performance: Epoch 24
\end{itemize}

\subsection{GAN (Basic)}

\textbf{Generator:} Similar to CNN architecture with 8 residual blocks

\textbf{Discriminator (PatchGAN):}
PatchGAN architecture~\cite{isola2017image} with the following structure:
\begin{itemize}
\item Conv 4×4, stride 2: 1 → 64 channels
\item LeakyReLU (0.2)
\item Conv 4×4, stride 2: 64 → 128 channels
\item LeakyReLU (0.2)
\item Conv 4×4, stride 2: 128 → 256 channels
\item LeakyReLU (0.2)
\item Conv 4×4, stride 1: 256 → 1 channel (score map)
\end{itemize}

\textbf{Training:}
\begin{itemize}
\item Adversarial loss: Binary cross-entropy with logits
\item Generator loss: $\mathcal{L}_G = \mathcal{L}_{L1} + 0.001 \times \mathcal{L}_{adv}$
\item Discriminator loss: $\mathcal{L}_D = 0.5 \times (\mathcal{L}_{real} + \mathcal{L}_{fake})$
\item Alternating updates: 1 discriminator step, 1 generator step
\item Epochs: 25
\item Training time: ~10 hours (24 min/epoch)
\item Best validation performance: Epoch 23
\end{itemize}

\subsection{GAN (Improved)}

\textbf{Generator Improvements:}
\begin{itemize}
\item 16 residual blocks (vs 8 in basic)
\item Spatial attention mechanism~\cite{zhang2019self} after residual blocks
\item Additional pre-exit convolution layer
\item Total parameters: ~0.6M
\end{itemize}

\textbf{Discriminator Improvements:}
\begin{itemize}
\item Deeper: 4 convolutional layers (vs 3)
\item Batch normalization after each conv (except first)
\item Additional 512-channel layer before output
\end{itemize}

\textbf{Training:}
\begin{itemize}
\item Feature matching loss added: $\mathcal{L}_{FM} = ||\phi(y) - \phi(G(x))||_1$
\item Generator loss: $\mathcal{L}_G = \mathcal{L}_{L1} + 0.01 \times \mathcal{L}_{adv} + 0.1 \times \mathcal{L}_{FM}$
\item Epochs: 10 (early stopping)
\item Training time: ~8 hours (48 min/epoch)
\item Best validation performance: Epoch 7
\end{itemize}

\subsection{DDPM (Denoising Diffusion Probabilistic Model)}

\textbf{Architecture:}
Conditional U-Net~\cite{ho2020denoising} with time embeddings for noise prediction.

\textbf{Diffusion Schedule:}
\begin{itemize}
\item Total timesteps $T = 100$ (vs standard 1000)
\item Linear noise schedule: $\beta_1 = 10^{-4}$ to $\beta_T = 0.02$
\item Forward process: $q(x_t | x_{t-1}) = \mathcal{N}(x_t; \sqrt{1-\beta_t} x_{t-1}, \beta_t I)$
\item Reverse process: $p_\theta(x_{t-1} | x_t, c) = \mathcal{N}(x_{t-1}; \mu_\theta(x_t, t, c), \Sigma_t)$
\end{itemize}

\textbf{Conditional U-Net:}
\begin{itemize}
\item Similar encoder-decoder structure to standard U-Net
\item Time embedding dimension: 256
\item Sinusoidal positional encoding for timesteps
\item Time embeddings injected into each residual block
\item Conditioning on adjacent slices (i-1, i+1) via concatenation
\item Self-attention at bottleneck layer
\item Total parameters: ~20.5M
\end{itemize}

\textbf{Training:}
\begin{itemize}
\item Objective: Predict noise $\epsilon$ at each timestep
\item Loss: $\mathcal{L} = ||\epsilon - \epsilon_\theta(x_t, t, c)||^2$
\item Optimizer: Adam, learning rate $10^{-4}$
\item Batch size: 8
\item Epochs: 20
\item Training time: ~6.3 hours (19 min/epoch)
\item Best validation loss: Epoch 16
\end{itemize}

\textbf{Sampling:}
\begin{itemize}
\item Start from Gaussian noise: $x_T \sim \mathcal{N}(0, I)$
\item Iteratively denoise for $t = T, T-1, ..., 1$
\item At each step: predict noise, compute mean, add noise (except t=0)
\item Total inference: 100 forward passes through 20.5M parameter network
\item Inference time: ~300ms per image (vs ~3-5ms for deterministic models)
\end{itemize}

\section{Extended Experimental Results}

\subsection{k=2 Formulation: Detailed Failure Analysis}

The k=2 formulation (using slices at positions i-2 and i+2 to predict slice i) represents a 6mm gap in the through-plane direction. At this spacing, four intermediate anatomical slices exist between the input slices.

\textbf{Why k=2 Failed:}
\begin{enumerate}
\item \textbf{Anatomical Discontinuity:} In prostate MRI at 1.5mm slice spacing, significant anatomical changes occur over 6mm. The prostate gland transitions from base to apex over ~40-50mm; thus 6mm represents 12-15\% of the organ's extent.

\item \textbf{Structural Transitions:} The intervening 6mm may span:
  \begin{itemize}
  \item Prostate zonal transitions (peripheral zone to central gland)
  \item Entry or exit of seminal vesicles
  \item Bladder neck to prostatic urethra transitions
  \item Prostatic capsule variations
  \end{itemize}

\item \textbf{Insufficient Boundary Information:} Even sophisticated deep learning models cannot reliably infer such complex intermediate anatomy from boundary conditions alone. The task becomes fundamentally ill-posed.

\item \textbf{Model Saturation:} All four architectures (CNN, U-Net, GANs) reached similar poor performance (SSIM ~0.56), suggesting the limitation arose from problem formulation rather than model capacity.
\end{enumerate}

\textbf{Training Dynamics with k=2:}

Despite extended training and hyperparameter exploration, k=2 models consistently plateaued:
\begin{itemize}
\item Training loss decreased normally
\item Validation metrics plateaued around epoch 15-20
\item No architecture achieved breakthrough performance
\item Increasing model capacity did not help
\end{itemize}

This suggested a fundamental ceiling imposed by task difficulty rather than optimization issues.

\subsection{Additional Qualitative Examples}

\begin{figure}[t]
\centering
\includegraphics[width=0.95\textwidth]{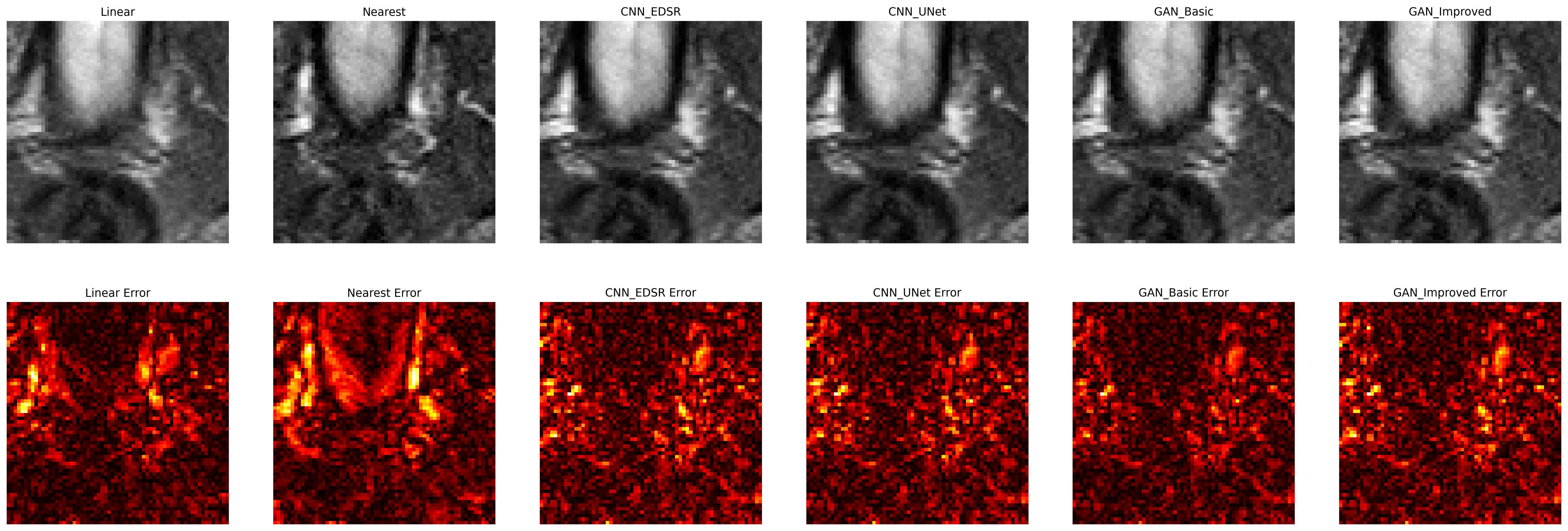}
\caption{Detailed zoomed comparison showing fine anatomical structures. Top row shows zoomed predictions from all methods. Bottom row shows corresponding error heatmaps. Deep learning methods preserve fine detail better than traditional interpolation.}
\label{fig:zoom}
\end{figure}

Figure~\ref{fig:zoom} provides zoomed views highlighting preservation of fine anatomical structures. U-Net and CNN methods maintain sharp boundaries at the prostatic capsule and gland-seminal vesicle interface. GANs show comparable performance with subtle texture differences. Traditional methods exhibit noticeable blurring.

\begin{figure}[t]
\centering
\includegraphics[width=0.95\textwidth]{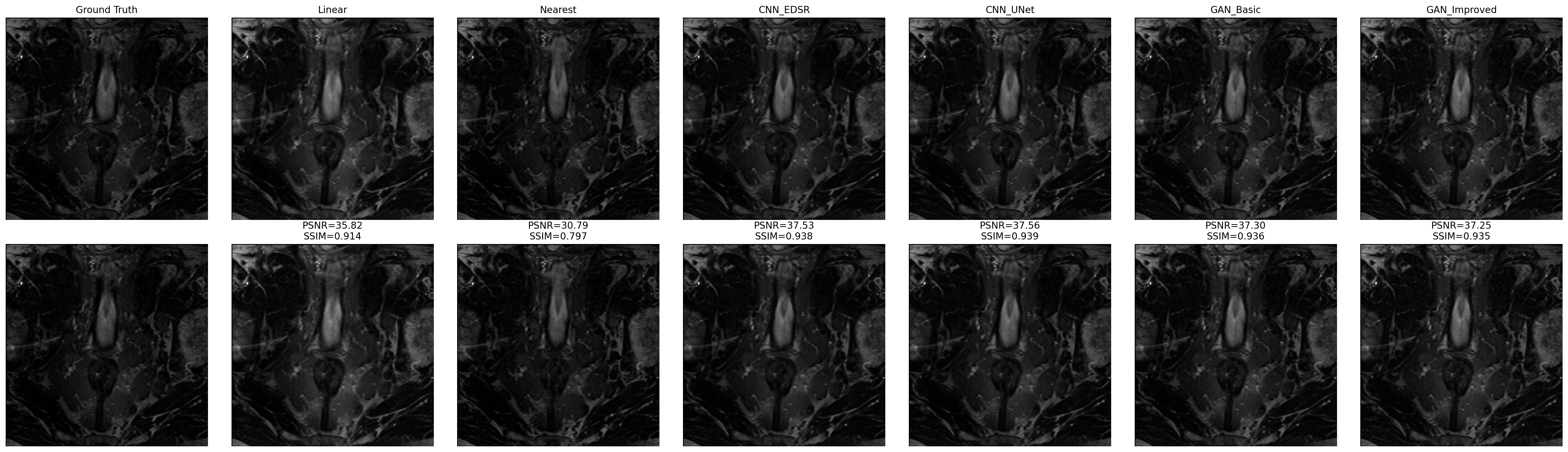}
\caption{Side-by-side comparison across two anatomical levels showing consistency of method performance. All deep learning methods generalize well across different anatomical regions.}
\label{fig:sidebyside}
\end{figure}

\subsection{Failure Cases and Limitations}

While deep learning methods substantially outperform traditional interpolation, certain challenging scenarios remain:

\textbf{Rapid Anatomical Transitions:}
At the prostatic apex and base, where anatomy changes rapidly between slices, all methods show increased error. These regions represent the boundaries of feasible interpolation even with k=1 formulation.

\textbf{Low Contrast Regions:}
In regions with minimal tissue contrast (e.g., within homogeneous central gland), interpolation accuracy depends heavily on intensity gradients. Subtle structural details may be smoothed.

\textbf{Motion Artifacts:}
When input slices contain patient motion between acquisitions, interpolation performance degrades as the assumption of anatomical continuity is violated.

\textbf{Edge Effects:}
At the very top and bottom of the prostate volume, where boundary slices are used, interpolation is inherently more difficult due to asymmetric context.

\section{Extended DDPM Analysis}

\subsection{Training Dynamics}

Despite DDPM's poor final performance, training proceeded stably:

\textbf{Training Loss Evolution:}
\begin{itemize}
\item Epoch 1: Train loss 0.1119, Val loss 0.0927
\item Epoch 5: Train loss 0.0956, Val loss 0.0861
\item Epoch 10: Train loss 0.0891, Val loss 0.0844
\item Epoch 16: Train loss 0.0808, Val loss 0.0813 (best)
\item Epoch 20: Train loss 0.0798, Val loss 0.0821
\end{itemize}

Loss decreased steadily with no overfitting (training and validation losses tracked closely). This indicates the model successfully learned the noise prediction task. The failure stems from task mismatch, not training issues.

\subsection{Why Standard Diffusion Objectives Fail for This Task}

DDPM optimizes for:
\begin{equation}
\mathcal{L} = \mathbb{E}_{t, x_0, \epsilon} [||\epsilon - \epsilon_\theta(x_t, t, c)||^2]
\end{equation}

This objective trains the model to predict noise, enabling stochastic sampling. However, slice interpolation requires:
\begin{itemize}
\item \textbf{Deterministic output:} Single ground truth slice must be reconstructed
\item \textbf{Pixel-accurate matching:} PSNR/SSIM metrics penalize any deviation
\item \textbf{Anatomical plausibility:} Generated structure must match actual anatomy
\end{itemize}

The stochastic sampling process introduces variability that degrades these metrics even if samples are perceptually reasonable.

\subsection{Alternative Approaches}

Several modifications might improve DDPM performance:

\textbf{DDIM Sampling:}
Denoising Diffusion Implicit Models (DDIM)~\cite{song2020denoising} enable deterministic sampling, potentially improving reconstruction metrics while maintaining diffusion model benefits.

\textbf{Increased Training:}
Standard DDPM implementations train for 500-1000 epochs with $T=1000$ timesteps. My 20 epochs and $T=100$ may have been insufficient.

\textbf{Hybrid Objectives:}
Combining noise prediction with direct reconstruction loss might bridge the gap between generation and reconstruction objectives.

\textbf{Conditional Guidance:}
Stronger conditioning on anatomical constraints (e.g., segmentation maps, structural priors) could constrain generation to anatomically valid outputs.

However, given the strong performance and 100× faster inference of deterministic models, these modifications appear academically interesting but practically unjustified for this specific task.

\begin{figure}[h]
\centering
\includegraphics[width=0.95\textwidth]{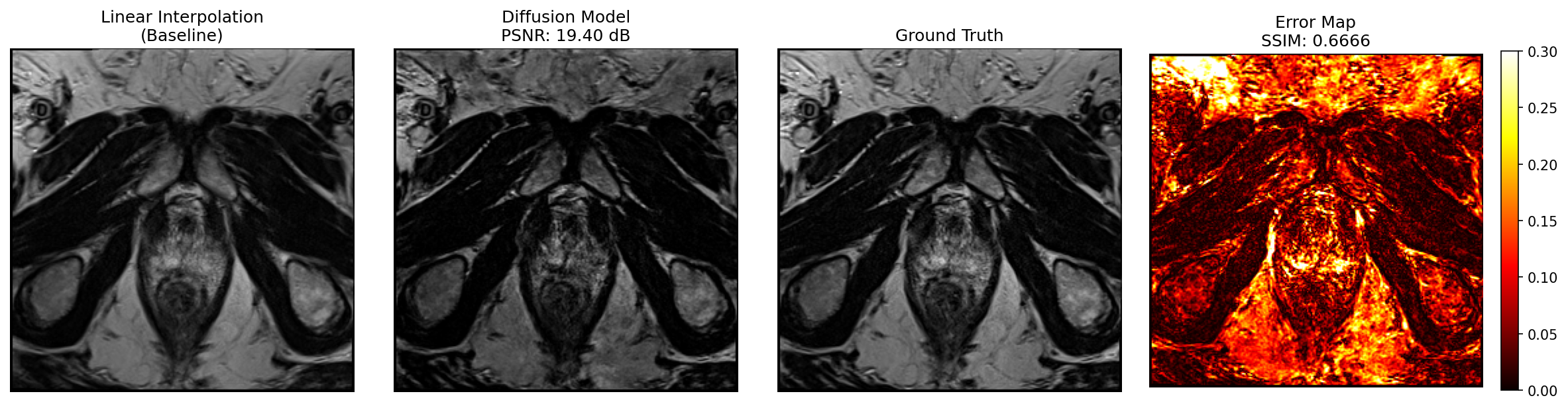}
\caption{Visual comparison of DDPM reconstruction failure. Left to right: Linear interpolation baseline, DDPM output (PSNR 19.40 dB), ground truth, error map (SSIM 0.6666). The error map reveals severe reconstruction failures (bright yellow/white regions) across anatomical boundaries, particularly in the prostate gland and surrounding structures. Despite stable training, the stochastic sampling process introduces artifacts that prevent accurate deterministic reconstruction.}
\label{fig:ddpm_failure}
\end{figure}

Figure~\ref{fig:ddpm_failure} provides visual evidence of DDPM's reconstruction failure. While the model successfully learned the noise prediction task (as evidenced by decreasing training loss), the stochastic sampling process produces outputs with severe anatomical inaccuracies. The error map shows concentrated failures (bright regions) at tissue boundaries and within the prostate gland, where precise reconstruction is most critical. Notably, the DDPM output appears somewhat smoother than linear interpolation but with incorrect anatomical details—a perceptually plausible but quantitatively inaccurate result that highlights the mismatch between generative and reconstructive objectives.

\subsection{Qualitative DDPM Failure Analysis}

\section{Limitations}

\textbf{Single Modality:} I focused exclusively on T2-weighted prostate MRI. Generalization to other sequences (T1, DWI), anatomical regions, or imaging modalities requires validation.

\textbf{Single Vendor/Field Strength:} All data came from 3T Siemens scanners. Performance on other vendors (GE, Philips) or field strengths (1.5T, 7T) is unknown.

\textbf{Resolution Assumptions:} My approach assumes consistent 1.5mm slice spacing. Clinical protocols vary (0.75-3mm), and thicker slices might exceed interpolation feasibility even with k=1 formulation.

\textbf{Evaluation Metrics:} PSNR and SSIM are imperfect proxies for clinical utility. Future work should include radiologist evaluation for diagnostic tasks (tumor detection, staging).

\textbf{Computational Resources:} My work required substantial GPU resources (hundreds of GPU-hours). This may limit reproducibility in resource-constrained settings, though inference is fast once models are trained.

\textbf{Patient Demographics:} The UCLA dataset may not represent all patient populations. Validation on diverse cohorts would strengthen generalizability claims.

\section{Clinical Implications and Future Directions}

\subsection{Clinical Potential}

Improved through-plane resolution enables:

\textbf{3D Reconstruction:} Higher isotropy improves 3D rendering for surgical planning and patient education, reducing interpolation artifacts in reconstructed volumes.

\textbf{Multiplanar Reformatting:} Clinicians often reformat axial acquisitions to sagittal or coronal views. Better through-plane resolution reduces reformatting artifacts and improves diagnostic confidence in non-native planes.

\textbf{Computer-Aided Diagnosis:} Many CAD systems assume isotropic data. Interpolation could enable deployment of such systems on clinical data without requiring acquisition protocol changes.

\textbf{Radiation Therapy Planning:} Improved resolution assists in precise tumor delineation for treatment planning.

\subsection{Deployment Considerations}

Clinical deployment requires:

\textbf{Regulatory Approval:} FDA clearance or equivalent for algorithm changes to clinical workflow.

\textbf{Clear Labeling:} Generated intermediate slices must be clearly marked to prevent misinterpretation as acquired data.

\textbf{Validation Studies:} Radiologist reader studies comparing diagnostic performance on original versus interpolated volumes.

\textbf{Integration:} Seamless integration into clinical PACS systems and radiology workflows.

\subsection{Future Research Directions}

\textbf{Multi-Sequence Integration:} Prostate MRI typically acquires multiple sequences (T2, T1, DWI, DCE). Leveraging information across sequences could improve interpolation quality, as different contrasts reveal complementary anatomical information.

\textbf{Uncertainty Quantification:} Providing confidence estimates for interpolated slices would help radiologists identify regions requiring scrutiny. Bayesian deep learning or ensemble methods could quantify prediction uncertainty.

\textbf{Active Learning:} In scenarios with limited high-resolution training data, active learning could identify most informative samples for annotation, reducing labeling burden.

\textbf{Task-Specific Optimization:} Rather than optimizing pixel-level metrics, training could target downstream tasks (tumor detection, segmentation) to ensure interpolation preserves clinically relevant information.

\textbf{Generalization Studies:} Evaluating performance across different scanners, institutions, and patient populations would assess real-world deployability.

\textbf{Real-Time Application:} Exploring deployment scenarios where interpolation occurs during or immediately after acquisition to provide higher resolution to radiologists without protocol changes.

\end{document}